\documentclass[12pt]{article}
\usepackage{fullpage,authblk,cite}
\usepackage{amsmath,amsthm,graphicx,amssymb,verbatim,listings}
\usepackage{wasysym,color,enumitem,mathcomp}
\usepackage[T1]{fontenc}     
\usepackage{lmodern, tipa}   
\usepackage[ pdftex, plainpages = false, pdfpagelabels,
                 pdfpagelayout = useoutlines,
                 bookmarks,
                 bookmarksopen = true,
                 bookmarksnumbered = true,
                 breaklinks = true,
                 linktocpage=all,
                 pagebackref=false,
                 colorlinks = true,
                 linkcolor = Blue,
                 urlcolor  = blue,
                 citecolor = BrickRed,
                 anchorcolor = green,
                 hyperindex = true,
                 hyperfigures
                 ]{hyperref}
\usepackage[usenames, dvipsnames]{xcolor}
\usepackage{xifthen} 

\newcommand{\tr}{\hbox{tr}}

\newcommand{\arxiv}[2][]{\ifthenelse{\isempty{#1}}{\href{http://arxiv.org/abs/#2}{{\tt arXiv:\allowbreak{}#2}}} {\href{http://arxiv.org/abs/#2}{{\tt arXiv:\allowbreak{}#2 [#1]}}}}

\newcommand{\booktitle}{\textsl}
\newcommand{\hrefdoi}[2]{\href{https://dx.doi.org/#1}{#2}}

\newcommand{\cP}{\mathcal{P}}

\newcommand{\cS}{\mathcal{S}}

\begin{document}
\title{QBism and the Ithaca Desiderata}
\author[$\dag$]{Blake C. Stacey}
\affil[$\dag$]{\small{\href{http://www.physics.umb.edu/Research/QBism/}{QBism
      Group}, Department of Physics, University of Massachusetts
    Boston, \par 100 Morrissey Boulevard, Boston~MA 02125, USA}}

\date{\small\today}

\maketitle

\begin{abstract}
In 1996, N.\ David Mermin proposed a set of desiderata for an
understanding of quantum mechanics, the ``Ithaca Interpretation''. In
2012, Mermin became a public advocate of QBism, an interpretation due
to Christopher Fuchs and R\"udiger Schack. Here, we evaluate QBism
with respect to the Ithaca Interpretation's six desiderata, in the
process also evaluating those desiderata themselves. This analysis
reveals a genuine distinction between QBism and the IIQM, but also a
natural progression from one to the other.
\end{abstract}

In 1996, N.\ David Mermin proposed the ``Ithaca Interpretation of
Quantum Mechanics''~\cite{Mermin:1996}. Rather than a complete story
about what quantum theory means, the IIQM was intended to be a set of
desiderata, a list of goals that a future interpretation of quantum
mechanics should meet in order to be considered satisfactory. In 2012,
Mermin became a public advocate of QBism, an interpretation due to
Christopher Fuchs and R\"udiger Schack~\cite{Mermin:2012, Fuchs:2013,
  VonBaeyer:2016}. This article evaluates QBism with respect to Mermin's
original six criteria. In four of the six cases, QBism qualifies
directly, and in the remaining two, a solid case can be made that
QBism satisfies the \emph{intuition behind} the desideratum. From the
viewpoint of QBism, each of the Ithaca desiderata puts a finger on a
legitimate and important question. What holds the IIQM back is its
insistence on allowing only a very limited kind of answer for some of
those questions.

QBism can be briefly defined as follows~\cite{DeBrota:2018b}:
\begin{quotation}
  \noindent An interpretation of quantum mechanics in which the ideas
  of \emph{agent} and \emph{experience} are fundamental. A ``quantum
  measurement'' is an act that an agent performs on the external
  world. A ``quantum state'' is an agent's encoding of her own
  personal expectations for what she might experience as a consequence
  of her actions. Moreover, each measurement outcome is a personal
  event, an experience specific to the agent who incites
  it. Subjective judgments thus comprise much of the quantum
  machinery, but the formalism of the theory establishes the standard
  to which agents should strive to hold their expectations, and that
  standard for the relations among beliefs is as objective as any
  other physical theory.
\end{quotation}
This article was prompted by a paper of de Ronde, Fern\'andez-Mouj\'an
and Massri which claimed that ``there is no interpretation more
distant from Mermin's original set of desiderata than
QBism''~\cite{DeRonde:2018}. As the discussion to follow will
hopefully make clear, this is untrue.

In what follows, we limit ourselves to brief quotations from Mermin's
prose, not just because Mermin's writing tends to make other
physicists' look flat and forgettable by comparison, but also because
for the present purposes, it is not necessary to establish that the
interpretation of quantum mechanics is a significant issue. We take it
as read that physicists \emph{should} be concerned with the topic ---
a claim perhaps more controversial than any specific choice of
interpretation. Thus, we proceed directly to Mermin's six desiderata
for an adequate interpretation,
which he summarizes as follows:
\begin{enumerate}
\item Is unambiguous about objective reality.
\item Uses no prior concept of measurement.
\item Applies to individual systems.
\item Applies to (small) isolated systems.
\item Satisfies generalized Einstein locality.
\item Rests on a prior concept of objective probability.
\end{enumerate}

After posing these criteria, the IIQM paper then presents two
technical theorems and discusses their philosophical implications. The
simplest approach is to address all of these topics in order.

\section{Objectivity}

First, quoting Mermin:
\begin{quotation}
\noindent A satisfactory interpretation should be unambiguous about
what has objective reality and what does not, and what is objectively
real should be cleanly separated from what is ``known''.
\end{quotation}
When confronted with the question of which elements of the standard
mathematical apparatus of quantum theory correspond to objective
reality and which do not, it is of course logically possible to reply,
``\emph{None} are fully objective --- all are at least tainted by the
subjective.'' This was the position of
E.\ T.\ Jaynes~\cite{Jaynes:1990}:
\begin{quotation}
  \noindent [O]ur present QM formalism is not purely epistemological;
  it is a peculiar mixture describing in part realities of Nature, in
  part incomplete human information about Nature --- all scrambled up
  by Heisenberg and Bohr into an omelette that nobody has seen how to
  unscramble. Yet we think that the unscrambling is a prerequisite for
  any further advance in basic physical theory. For, if we cannot
  separate the subjective and objective aspects of the formalism, we
  cannot know what we are talking about; it is just that simple.
\end{quotation}
Jaynes believed that to unscramble the Heisenberg--Bohr omelette, ``we
need to find a different formalism, isomorphic in some sense but based
on different variables''~\cite{Jaynes:1989}.

QBism is unambiguous about what is ``known'' (or, to put it better,
\emph{believed}) and what is objective. In particular, quantum states
belong on the subjective side of the line, while for any individual
agent, her personal experiences are empirical and
incontrovertible~\cite{Mermin:2018}. Moreover, in QBism the quantum
formalism itself encodes a \emph{normative standard} that any agent
should strive to attain. This standard is objective, even though the
expectations held by any individual agent are their own subjective
belongings. But making this clear requires a Jaynesian unscrambling:
The textbook formalism, good as it is for so many things, isn't the
best suited for resolving this question. (This should be no
surprise. The idea of a single picture of a theory that works equally
well for all problems is antithetical to a physicist's training and
lifestyle.) Fuchs and colleagues have identified a certain
probabilistic representation of the Born Rule as the cleanest
statement of the objective normative standard that quantum theory
expresses~\cite{Fuchs:2017}.

Mermin goes on to say, ``Indeed, knowledge should not enter at a
fundamental level at all.'' In QBism, the term \emph{knowledge} is
deprecated, and \emph{belief} or \emph{expectation} are preferred: The
latter terminology carries less of a connotation that different agents
must necessarily come into agreement, as a matter of principle (though
they may often do so in practice). According to QBism,
\emph{expectation} is a fundamental part of quantum theory, which is
different from expectation (or ``information'' or ``knowledge'') being
a fundamental ingredient of reality itself. Before there were agents,
there was reality, but there were no expectations.

A useful comparison can be made to special relativity. Consider
Einstein's postulates, and the dramatic tension between them: Inertial
observers can come to agree upon the laws of physics, but they cannot
agree upon a standard of rest. These axioms are conveniently expressed
in terms of what agents can and cannot do, yet they are more than
``mere'' engineering, because they apply to all agents. Or, to say it
another way, any agent should take heed of the theory when trying to
realize their own aspirations. In QBism, the role of quantum theory is
analogous.

Most of Mermin's writings on QBism have addressed how it gives meaning
to the mathematical formalism of quantum physics. Other QBist authors
have put more emphasis on QBism \emph{as a project,} that is, as a
motivator for technical research~\cite{Fuchs:2017, Fuchs:2013b}. (I
myself fall into this latter group~\cite{DeBrota:2018b, Fuchs:2016,
  Fuchs:2016b, DeBrota:2018, Stacey:2018}, since I have not yet
attained that stage of a physicist's career where one can safely write
papers that lack equations. Quantum foundations, properly understood,
is not just about pretty words, but a matter of Sylow subgroups,
Galois fields and integral octonions~\cite{Zhu:2010, Bengtsson:2017,
  Appleby:2017b, Stacey:2017, Stacey:2016b}.) What features of the
natural world make quantum theory a good calculus of expectations?
What internally consistent alternatives to quantum theory can be
imagined, and how do they illuminate the particularities of quantum
theory itself? In short, to quote the title Mermin gave to his second
paper on the IIQM~\cite{Mermin:1998}, ``What is quantum mechanics
trying to tell us?'' The project of QBism aims to answer exactly that,
and the first step is to ``be unambiguous about what has objective
reality and what does not''.

\section{Measurement}
On the second desideratum, quoting Mermin again:
\begin{quotation}
\noindent The view that physics can offer nothing more than an
algorithm telling you how to get from a state preparation to the
results of a measurement seems to me absurdly anthropocentric; so does
limiting what we can observe to what we can produce (``state
preparation'' being one of the things you can do with a ``measurement
apparatus''). Physics ought to describe the unobserved unprepared
world. ``We'' shouldn't have to be there at all.
\end{quotation}
The QBist answer is twofold. First, QBism discards the limitation on
``what we can observe'' --- any action by an agent, interacting with
the outside world, is in principle a quantum measurement. In his
second essay on the IIQM, Mermin argues that ``the very much broader
concept of correlation ought to replace measurement in a serious
formulation of what quantum mechanics is all
about''~\cite{Mermin:1998}. QBism does the job more directly by
broadening the concept of \emph{measurement} itself. Any action taken
by an agent upon the external world is, in principle, a quantum
measurement, and any experience incited by an action is a measurement
outcome. The supposed limitation of quantum theory's validity to
bench-top laboratory procedures simply does not exist. Second, the
fact that the theory resists cutting the agent out of it is a
statement about the character of the physical world. ``We'' don't have
to be here, but the fact that this particular theory is helpful to us
now that we are --- \emph{that} says something about nature.

This item is the point of greatest divergence between QBism and the
IIQM. (Mermin quipped in 2012, ``Like Barack Obama's view of marriage,
my thinking about quantum foundations has
evolved''~\cite{Mermin:2012b}.) From the QBist perspective, the IIQM
reaches for objectivity too soon, thereby missing its chance to hear
what the theory is trying to say.

\section{Individual and Isolated Systems}
Mermin states, regarding his third criterion,
\begin{quotation}
\noindent The theory should describe individual systems --- not just
ensembles.
\end{quotation}
QBism is fine with individual systems, because it adopts a school of
probability theory intended for single-shot situations. This is a
healthy thing to do, even in classical science~\cite{Baez:2003,
  Appleby:2005a, Appleby:2005b, Diaconis:2018}.

Mermin elaborates as follows upon his fourth criterion:
\begin{quotation}
\noindent The theory should describe small isolated systems without
having to invoke interactions with anything external. [\,\ldots] In
particular I would like to have a quantum mechanics that does not
require the existence of a ``classical domain''.  Nor should it rely
on quantum gravity, or radiation escaping to infinity, or interactions
with an external environment for its \emph{conceptual} validity. These
complications may be important for the practical matter of explaining
why certain probabilities one expects to be tiny are, in fact
tiny. But it ought to be possible to deal with high precision and no
conceptual murkiness with small parts of the universe if they are to
high precision, isolated from the rest.
\end{quotation}
QBism nowhere demands the existence of a ``classical domain'', nor
does it rely on the other kinds of dodges like radiation leaking away
to infinity. It is of course \emph{capable} of treating the time
evolution of an open quantum system as it is at handling any
application of the standard quantum formalism.

Mermin states his fifth criterion, an Einsteinian notion of locality,
as follows.
\begin{quotation}
\noindent Objectively real internal properties of an isolated
individual system should not change when something is done to another
non-interacting system.
\end{quotation}
Fuchs, in particular, has used exactly this argument to make the case
that quantum states are subjective. The argument is essentially the
reason Einstein gave for why quantum states cannot be intrinsic
``physical conditions'' of systems~\cite{Einstein:1950}. Einstein wrote,
\begin{quotation}
  \noindent Consider a mechanical system constituted of two partial
  systems $A$ and $B$ which have interaction with each other only
  during limited time. Let the $\psi$ function before their
  interaction be given. Then the Schr\"odinger equation will furnish
  the $\psi$ function after their interaction has taken place. Let us
  now determine the physical condition of the partial system $A$ as
  completely as possible by measurements. Then the quantum mechanics
  allows us to determine the $\psi$ function of the partial system $B$
  from the measurements made, and from the $\psi$ function of the
  total system. This determination, however, gives a result which
  depends upon \emph{which} of the determining magnitudes specifying
  the condition of $A$ has been measured (for instance coordinates
  \emph{or} momenta). Since there can be only \emph{one} physical
  condition of $B$ after the interaction and which can reasonably not
  be considered as dependent on the particular measurement we perform
  on the system $A$ separated from $B$ it may be concluded that the
  $\psi$ function is not unambiguously coordinated with the physical
  condition. This coordination of several $\psi$ functions with the
  same physical condition of system $B$ shows again that the $\psi$
  function cannot be interpreted as a (complete) description of a
  physical condition of a unit system.
\end{quotation}
The QBist take on this simply replaces Einstein's notion of
probability for a personalist Bayesian one, and lets go of the desire
to complete the description using hidden variables. (For further
discussion, see~\cite{Fuchs:2016, Fuchs:2002}.)  Plainly, QBism and
the IIQM agree about Einstein locality.

\section{Objective Probability versus Objective Indeterminism}
Mermin gives his sixth desideratum as the following:
\begin{quotation}
\noindent It suffices (\emph{for now}) to base the interpretation of
quantum mechanics on the (\emph{yet to be supplied}) interpretation of
objective probability.
\end{quotation}
In other words, ``objective probability'' is a rug under which as many
problems as possible are to be swept --- a rug whose existence is not
known, but hoped for. According to QBism, no satisfactory
interpretation of objective probability will ever be supplied. So,
QBism is definitely at odds with this desideratum. But Mermin himself
began to doubt that ``objective probability'' or ``propensity'' could
be made a sensible idea, long before he adopted QBism. As he wrote to
Fuchs in January 2006~\cite{Fuchs:2014},
\begin{quotation}
\noindent You persuaded me quite soon that ``objective probability'' was
problematic.   Until I met you I had never taken the notion of
subjective probability seriously, or even known very much about it.
While I'm still not convinced (sorry) that you've got it right
either, I'm much more aware that one of the pillars of the IIQM is
much more fragile than I thought.
\end{quotation}
In QBism, the numerical value of any probability is the personal
property of the agent who assigns it. However, the QBist world is one
of \emph{objective, irreducible indeterminism.} Physical reality so
overflows with richness that no notion of ``propensity'' can ever be
adequate. (For further discussion, see \cite[\S 2.2]{Fuchs:2017} and
\cite[footnote 6]{Fuchs:2017b}.) The character of that deep and
thoroughgoing indeterminism implies conditions which any agent who
uses probabilities should strive to meet; we call those conditions
quantum theory. Those looking for objectivity in the numerical values
of probabilities are just looking in the wrong
place~\cite{Fuchs:2016}.

\section{Theorems}

After laying out the six Ithaca desiderata, Mermin discusses two
technical theorems. The first, which concerns the mulitplicity of
pure-state decompositions of a general mixed state, is often called
the HJW theorem after Hughston, Jozsa and
Wootters~\cite{Hughston:1993}. (The result has a prior history going
through Ochs~\cite{Ochs:1981}, Jaynes~\cite{Jaynes:1957} and
Schr\"odinger~\cite{Schroedinger:1936}, and caused at least one
statistical physicist to take strong issue with von
Neumann~\cite{Stacey:2016}.) Mermin uses this and Einstein locality to
argue that different decompositions of the same mixed state must be
physically equivalent:
\begin{quotation}
\noindent If you take Desideratum (5) seriously, then there can be no more
objective reality to the different possible realizations of a density
matrix, than there is to the different possible ways of expanding a
pure state in terms of different complete orthonormal sets. [\,\ldots]
In the case of an individual system, the density matrix must be a
fundamental and irreducible objective property, whether or not it is a
pure state.
\end{quotation}
We have here a good example of how QBism echoes the IIQM while at the
same time contrasting with it. Mermin's argument here becomes a
statement entirely in line with QBism when one performs the admittedly
radical move of dropping the term ``objective property''. In QBism,
any quantum state, mixed or pure, is simply a catalogue of an agent's
expectations about what that agent might experience as a result of
interacting with an external system. Consequently, just as Mermin was
aiming for, there is no category distinction in QBism between rank-1
projectors and other density operators. The former simply enjoy the
additional property of being extremal in the set of valid catalogues
of expectation. Both QBism and the IIQM treat all quantum states on
the same footing, but they differ on what that footing must be.

The technical program pursued by Fuchs and colleagues underscores this
point, emphasizing that any quantum state on a $d$-dimensional Hilbert
space can be represented as a probability distribution over the
outcomes of a reference measurement~\cite{Fuchs:2013b, Fuchs:2016,
  Fuchs:2016b, DeBrota:2018, Appleby:2017}. Thus, pure states are not
categorically different from mixed states, any more than, say,
Gaussian curves are from other probability distributions. In the
simplest case, that of a single qubit, an appropriate choice of
reference measurement reveals that quantum state space (i.e., the
Bloch ball) is isomorphic to the set $\cP$ of four-element probability
vectors that satisfies
\begin{equation}
 \frac{1}{6} \leq \sum_{i=1}^4 p(i) p'(i)
\leq \frac{1}{3},\ \forall\ p,p' \in \cP.
\end{equation}
Pure qubit states are exactly those that saturate the upper bound on
their Euclidean norm (which, equivalently, is a lower bound on their
R\'enyi 2-entropy). The constraints become more complicated in higher
dimensions, but the principle remains the same~\cite{Fuchs:2013b,
  Fuchs:2016, Fuchs:2016b}.

Thus, we can observe a general prejudice in the community about which
of the Ithaca desiderata are the most valued. Interpretations of
quantum mechanics often rush to excise ``measurement'' from their
vocabulary, but seldom to accord equal status to pure and mixed
states.

Mermin states the second theorem as follows.
\begin{quotation}
\noindent Given a system $\cS = \cS_1 \oplus \cS_2$ with density
matrix $W$, then $W$ is completely determined by the values of $\tr W
A \otimes B$ for an appropriate set of observable pairs $A, B$, where
$A = A \otimes 1$ is an observable of subsystem $\cS_1$ and $B = 1
\otimes B$ is an observable of subsystem $\cS_2$.
\end{quotation}
As Mermin shows, this is provable from the standard formalism of
quantum mechanics~\cite{Beria:1980, Wootters:1990}. Since QBism is an
interpretation of quantum mechanics, a way of investing the
mathematics with meaning, this theorem is of course consistent with
it. The result that, in Mermin's phrasing, ``the correlations among
all the subsystems completely determine the density matrix for the
composite system they make up'' has come to be known as
\emph{tomographic locality.} Several approaches to reconstructing
quantum theory from operational principles have invoked this as a
postulate; there, its main role is to distinguish the orthodox quantum
theory, with its complex Hilbert spaces, from its ``foil theories''
defined over real and quaternionic
algebras~\cite{Hardy:2012}. However, tomographic locality is not the
only postulate that has been used in that role~\cite{Stacey:2018,
  Caves:2002, Barnum:2014, Garner:2017}; nor has it played a major
role in the QBist effort to reconstruct quantum
theory~\cite{Fuchs:2016b}. For further discussion of related
technicalities, see \cite[\S 5]{Fuchs:2002} and \cite{Barnum:2010}, as
well as \cite[pp.\ 217--52, 499--500]{Fuchs:2011}.

In the context of his original Ithaca desiderata, Mermin takes the
meaning of the second theorem to be that ``the fundamental irreducible
objective character of an individual system is entirely specified by
all the correlations among any particular set of the subsystems into
which it can be decomposed.'' The slogan of the IIQM was
\emph{correlations without correlata}~\cite{Mermin:1998}. Later,
Mermin wrote to Fuchs about difficulties with this
idea~\cite{Fuchs:2014}.
\begin{quotation}
\noindent Not unrelated to [the problem with ``objective
  probability''], the notion of ``correlation'' is not well defined,
beyond my assertion that it means nothing more than ``joint
distribution''. But what does it mean to say that joint distributions
are fundamental, while conditional distributions, which can be
constructed from joints, have no physical meaning? And what are these
joint distributions describing?
\end{quotation}
In an interview with Max Schlosshauer~\cite{Mermin:2013}, Mermin said,
\begin{quotation}
  \noindent What led me to stop giving physics colloquia on the IIQM
  after only a year was the obvious question: ``Correlations between
  what?''  Abner Shimony aptly complained that the Ithaca
  Interpretation ``had no foreign policy.''
\end{quotation}

Another reason why the technical side of QBism has placed less
emphasis on these two theorems than the IIQM did is that they are not
strictly specific to quantum mechanics. The Spekkens toy model, a
theory defined explicitly in terms of local hidden variables, has both
the multiplicity of mixed-state decompositions and tomographic
locality~\cite{Spekkens:2007}. If the slogan of the IIQM was
\emph{correlations without correlata,} then the motto of the Spekkens
toy model could be \emph{correlations with concealed correlata}: The
model has local hidden variables, but the observer is constrained from
ever having full knowledge about what values those variables take. The
Spekkens toy model reproduces many things first discovered in quantum
theory, such as teleportation and the no-cloning theorem, indicating
that however intriguing and useful these features might be in
particular applications, they are not where the \emph{essence} of
quantum theory lies. To find that, we have to dig
deeper. Consequently, the theorems that the IIQM had centered appear
more peripheral --- not insignificant, but secondary, to be derived
rather than assumed~\cite{Fuchs:2016b}.

\section{Conclusions}

In summary, QBism meets four of the six Ithaca desiderata in a
straightforward way, and in a more subtle fashion, it addresses
the underlying concerns of the other two. Of the two theorems
emphasized by Mermin's original IIQM paper, QBism readily synergizes
with the first and has no quarrel with the second. The technical
research done under the QBist banner has placed less emphasis on that
second theorem than the IIQM papers did, partly because other
mathematical statements seem capable of standing in the same place.

The paper that prompted this commentary characterized QBism as ``a
neo-Bohrian, anti-realist and instrumentalist account'' of quantum
mechanics~\cite{DeRonde:2018}. Being both \emph{neo-Bohrian} and
\emph{anti-realist} would be a neat trick, since Bohr was a
realist~\cite{Bohr:1939, Folse:1985, Barad:2007}. QBism is also
realist --- it is simply not \emph{na\"\i{}vely realist} about any
particular mathematical element of the quantum formalism. It and
Bohrian thought are two detectably distinct examples of
\emph{participatory realism}~\cite{Fuchs:2017b, Cabello:2017}. In
fact, QBism has drawn as much upon the thinking of \emph{Einstein} as
upon Bohr~\cite{Stacey:2018b}, a point that our discussion of the
locality desideratum suggested. As Fuchs declared~\cite{Fuchs:2014},
``When Einstein was right, he was really right!'' QBism comes down in
places closer to Pauli than to Bohr, while remaining not completely
satisfied with \emph{any} of the founders~\cite{Fuchs:2017}.

Whether ``instrumentalism'' has any meaning other than as a
philosopher's swear word, we leave as an exercise to the interested
reader. It seems, though, that characterizing QBism as
``instrumentalist'' misses the animating force behind the interpretive
effort: to take the question ``What is quantum mechanics trying to
tell us?''\ as the impetus for new physics. Directly identifying the
surface-level elements of quantum theory's mathematical formalism with
physical reality is an easy path to start down, but one that sooner,
rather than later, leads to confusion. The sought-after clarity
recedes, leaving new vaguenesses to stand beside the old. The ethos of
QBism is to resist the lure of easy answers. As Schr\"odinger once
wrote~\cite{Schroedinger:1951},
\begin{quotation}
  \noindent In an honest search for knowledge you quite often have to
  abide by ignorance for an indefinite period. Instead of filling a
  gap by guesswork, genuine science prefers to put up with it; and
  this, not so much from conscientous scruples about telling lies, as
  from the consideration that, however irksome the gap may be, its
  obliteration by a fake removes the urge to seek after a tenable
  answer. So efficiently may attention be diverted that the answer is
  missed even when, by good luck, it comes close at hand. The
  steadfastness in standing up to a \emph{non liquet,} nay in
  appreciating it as a stimulus and a signpost to further quest, is a
  natural and indispensable disposition in the mind of a scientist. \qed
\end{quotation}

\bigskip

This research was supported by the John Templeton Foundation. The
opinions expressed in this publication are those of the author and do
not necessarily reflect the views of the John Templeton Foundation.

\end{document}